\documentclass[preprint,dvipdfmx]{ptephy}

\preprintnumber{KOBE-TH-20-07, KYUSHU-HET-219}

\usepackage{amssymb}
\usepackage{amsthm}
\usepackage{amsmath}
\usepackage{booktabs}
\usepackage{bbm}
\usepackage{bm}
\usepackage{mathtools}
\usepackage{simplewick}
\usepackage{subcaption}

\allowdisplaybreaks

\DeclareMathOperator{\tr}{tr}

\let\Re\relax
\DeclareMathOperator{\Re}{Re}

\numberwithin{equation}{section}

\begin{document}

\title{Gradient flow exact renormalization group}

\author{%
\name{\fname{Hidenori} \surname{Sonoda}}{1} and
\name{\fname{Hiroshi} \surname{Suzuki}}{2,\ast}
}

\address{%
\affil{1}{Physics Department, Kobe University, Kobe 657-8501, Japan}
\affil{2}{Department of Physics, Kyushu University, 744 Motooka, Nishi-ku,
Fukuoka 819-0395, Japan}
\email{hsuzuki@phys.kyushu-u.ac.jp}
}

\date{\today}

\begin{abstract}
The gradient flow bears a close resemblance to the coarse graining, the guiding
principle of the renormalization group (RG). In the case of scalar field
theory, a precise connection has been made between the gradient flow and the RG
flow of the Wilson action in the exact renormalization group (ERG) formalism.
By imitating the structure of this connection, we propose an ERG differential
equation that preserves manifest gauge invariance in Yang--Mills theory. Our
construction in continuum theory can be extended to lattice gauge theory.
\end{abstract}

\subjectindex{B32, B05}
\maketitle

\section{Introduction}
\label{sec:1}
The gradient flow~\cite{Narayanan:2006rf,Luscher:2009eq,Luscher:2010iy,%
Luscher:2011bx,Luscher:2013cpa,Luscher:2013vga} is a continuous deformation of
a gauge field configuration~$A_\mu^a(x)$ along a fictitious time~$t\geq0$. It
is given by a gauge-covariant diffusion equation
\begin{equation}
   \partial_t B_\mu^a(t,x)=D_\nu G_{\nu\mu}^a(t,x),\qquad
   B_\mu^a(t=0,x)=A_\mu^a(x),
\label{eq:(1.1)}
\end{equation}
where
\begin{equation}
   G_{\mu\nu}^a(t,x)
   \equiv\partial_\mu B_\nu^a(t,x)-\partial_\nu B_\mu^a(t,x)
   +f^{abc}B_\mu^b(t,x)B_\nu^c(t,x)
\label{eq:(1.2)}
\end{equation}
is the field strength of the flowed or diffused
field~$B_\mu^a(t,x)$,\footnote{$f^{abc}$ is the structure constant defined from
the anti-hermitian generator~$T^a$ of the gauge group
by~$[T^a,T^b]=f^{abc}T^c$.} and
\begin{equation}
   D_\mu X^a(t,x)\equiv\partial_\mu X^a(t,x)+f^{abc}B_\mu^b(t,x)X^c(t,x)
\label{eq:(1.3)}
\end{equation}
is the covariant derivative with respect to~$B_\mu^a(t,x)$. The gradient flow
bears a close resemblance to the coarse graining along renormalization group
(RG) flows~\cite{Wilson:1973jj}. This aspect of the gradient flow has been
investigated from various perspectives~\cite{Luscher:2013vga,Kagimura:2015via,%
Yamamura:2015kva,Aoki:2016ohw,Pawlowski:2017rhn,Makino:2018rys,Abe:2018zdc,%
Carosso:2018bmz,Carosso:2018rep,Sonoda:2019ibh,Carosso:2019qpb,%
Matsumoto:2020lha}. In this paper we further our understanding of how the
gradient flows are related to the RG flows by using the exact renormalization
group (ERG) formalism (for reviews of ERG, see for
instance~Refs.~\cite{Pawlowski:2005xe,Igarashi:2009tj,Rosten:2010vm}).

In scalar field theory, the analogue of~Eq.~\eqref{eq:(1.1)} would
be~\cite{Capponi:2015ucc}
\begin{equation}
   \partial_t\varphi(t,x)=\partial_\mu\partial_\mu\varphi(t,x),\qquad
   \varphi(t=0,x)=\phi(x).
\label{eq:(1.4)}
\end{equation}
It is actually possible to make a precise connection between the gradient flow
and the flow of a Wilson action under ERG~\cite{Sonoda:2019ibh} (see
also~Ref.~\cite{Matsumoto:2020lha}). In $D$-dimensional Euclidean space, the
ERG differential equation for the Wilson action~$S_\tau[\phi]$ (the so-called
Wilson--Polchinski equation~\cite{Polchinski:1983gv}) reads, in terms of
\emph{dimensionless} variables,\footnote{Throughout this paper, we use
abbreviations,
\begin{equation}
   \int_p\equiv\int\frac{d^Dp}{(2\pi)^D},\qquad
   \delta(p)\equiv(2\pi)^D\delta^{(D)}(p).
\label{eq:(1.5)}
\end{equation}
}
\begin{align}
   \frac{\partial}{\partial\tau}\,
   e^{S_\tau[\phi]}
   &=\int_p\biggl(
   \left\{
   \left[\frac{\Delta(p)}{K(p)}+\frac{D+2}{2}
   -\frac{\eta_\tau}{2}\right]
   \phi(p)
   +p_\mu\frac{\partial}{\partial p_\mu}\phi(p)
   \right\}
   \frac{\delta}{\delta\phi(p)}
\notag\\
   &\qquad\qquad{}
   +\frac{1}{p^2}\left[
   2\frac{\Delta(p)}{K(p)}k(p)
   +2p^2\frac{dk(p)}{dp^2}
   -\eta_\tau k(p)
   \right]
   \frac{1}{2}
   \frac{\delta^2}{\delta\phi(p)\delta\phi(-p)}
   \biggr)\,
   e^{S_\tau[\phi]},
\label{eq:(1.6)}
\end{align}
where $K$ and $k$ are cutoff functions satisfying
\begin{equation}
   K(p)=\begin{cases}
   1&\text{for $|p|\to0$},\\
   0&\text{for $|p|\to\infty$},\\
   \end{cases},\qquad
   k(p)\stackrel{|p|\to0}{\to}0,
\label{eq:(1.7)}
\end{equation}
and
\begin{equation}
   \Delta(p)\equiv-2p^2\frac{dK(p)}{dp^2}.
\label{eq:(1.8)}
\end{equation}
The origin of the anomalous dimension~$\eta_\tau$ in the above has been
elucidated in~Ref.~\cite{Igarashi:2016qdr}. Particularly for~$K(p)=e^{-p^2}$, it
has been shown~\cite{Sonoda:2019ibh} that the correlation functions of the
diffused field~$\varphi(t,x)$, defined by~Eq.~\eqref{eq:(1.4)}, calculated with
the ``bare'' action~$S_{\tau=0}[\phi]$ are essentially identical to the
correlation functions of the elementary field~$\phi(x)$ calculated with the
Wilson action~$S_\tau[\phi]$;\footnote{In~Ref.~\cite{Sonoda:2019ibh}, a
particular choice~$k(p)=K(p)\left(1-K(p)\right)$~\cite{Polchinski:1983gv} has
been made, but this restriction can be relaxed; see below.} the flow time~$t$
in~Eq.~\eqref{eq:(1.4)} and the scale parameter~$\tau$ in the ERG
equation~\eqref{eq:(1.6)} are related by~$t=e^{2\tau}-1$. We will review this
observation in the next section. The connection between the gradient flow and
ERG can naturally explain~\cite{Sonoda:2019ibh} why the local products of the
diffused field remain finite under the wave function renormalization of
elementary fields~\cite{Luscher:2011bx,Luscher:2013cpa}: we first obtain the
Wilson action~$S_\tau[\phi]$ by integrating over field modes whose momenta are
higher than a cutoff (corresponding to the parameter~$\tau$), and then the
correlation functions of the field~$\phi(x)$ are obtained by integration of the
field-modes with momenta less than the cutoff, and thus are
finite.\footnote{The argument given for scalar field theory
in~Ref.~\cite{Sonoda:2019ibh} assumes the same flow time for the diffused
fields (because the flow time is identified with the scale parameter in the
Wilson action), but it somewhat extends the result
of~Refs.~\cite{Luscher:2011bx,Luscher:2013cpa} for gauge theory, in that it
applies not only to the continuum limit around the Gaussian fixed point but
also to that around a non-trivial fixed point such as the Wilson--Fisher fixed
point.}

It is of great interest to find such a connection between the gradient and ERG
flows in gauge theory; it would provide a natural understanding of the
finiteness of the correlation functions of the diffused gauge
field~\eqref{eq:(1.1)} in the continuum limit~\cite{Luscher:2011bx} (see
also~Ref.~\cite{Hieda:2016xpq}). The manifest gauge covariance of the gradient
flow~\eqref{eq:(1.1)} would suggest a manifestly gauge-invariant ERG
formulation of gauge theory. It appears quite difficult, however, to make such
a direct connection. The gradient flow equation in gauge
theory~\eqref{eq:(1.1)} is highly non-linear compared with the flow
equation~\eqref{eq:(1.4)} in scalar field theory, which is linear and solvable.
The argument of~Ref.~\cite{Sonoda:2019ibh} took advantage of this simplicity.

In this paper, we look at the problem from a different perspective. We first
derive, on the basis of the result of~Ref.~\cite{Sonoda:2019ibh}, a
representation of the Wilson action~$S_\tau[\phi]$ directly in terms of the
diffused field~$\varphi(t,x)$ in~Eq.~\eqref{eq:(1.4)}. We can readily
generalize this representation to the Yang--Mills theory, simply by replacing
$\varphi(t,x)$ by the diffused gauge field~$B_\mu^a(t,x)$
in~Eq.~\eqref{eq:(1.1)}. We regard this as a \emph{definition\/} of the Wilson
action. We will argue that our construction of the Wilson action effectively
implements an ultraviolet cutoff in~$S_\tau[A]$. From the representation
of~$S_\tau[A]$, we see that $S_\tau[A]$ and~$S_{\tau=0}[A]$ give identical
partition functions. The corresponding ERG transformation thus preserves the
partition function, as is usually required for ERG. We can also see that
$S_\tau[A]$ possesses manifest gauge invariance as long as the initial
action~$S_{\tau=0}[A]$ is gauge invariant; the ERG thus preserves gauge
invariance. We then derive an ERG differential equation by taking the
$\tau$~derivative of~$S_\tau[A]$. The resulting ERG equation is written
entirely in terms of~$S_\tau[A]$, and once this ERG equation is obtained, we
may forget about the original representation of~$S_\tau[A]$ based on the
gradient flow.

This paper is organized as follows. In~Sect.~\ref{sec:2}, we review the
argument of~Ref.~\cite{Sonoda:2019ibh} and derive a representation,
Eq.~\eqref{eq:(2.13)}, of the Wilson action in terms of the diffused field;
this representation becomes the basis of our construction of the Wilson
action~$S_\tau[A]$ in Yang--Mills theory in~Sect.~\ref{sec:3.1}. We analyze the
gauge invariance of~$S_\tau[A]$ in~Sect.~\ref{sec:3.2}; we show that $S_\tau[A]$
possesses manifest gauge invariance as long as the initial
action~$S_{\tau=0}[A]$ is gauge invariant. This implies that the ERG
differential equation, Eq.~\eqref{eq:(3.25)}, that we derive
in~Sect.~\ref{sec:3.3} preserves gauge invariance. In~Sect.~\ref{sec:3.4}, we
solve the ERG equation in the lowest approximation, i.e., in the lowest order
in a parameter~$\lambda$~\eqref{eq:(3.10)}. This parameter turns out to provide
a convenient expansion parameter analogous to the conventional gauge coupling.
In~Sect.~\ref{sec:4}, we generalize the construction of the Wilson action
in~Sect.~\ref{sec:3.1} to lattice gauge theory. We conclude the paper
in~Sect.~\ref{sec:5}. There is a short appendix to~Sect.~\ref{sec:3} about the
normalization of the gauge field.

In this paper, we only present the basic idea and basic equations for our
formulation of Yang--Mills theory; we defer possible applications for future
studies.

\section{Scalar field theory}
\label{sec:2}
As pointed out in~Ref.~\cite{Sonoda:2015bla}, the change of a Wilson
action~$S_\tau$ under a change of the cutoff scale in~Eq.~\eqref{eq:(1.6)} can
be formulated as an equality of \emph{modified\/} correlation functions. In
terms of \emph{dimensionless\/} variables, Eq.~(38)
of~Ref.~\cite{Sonoda:2015bla} with $t\to0$, $\Delta t\to\tau$,
and~$e^{\Delta t\gamma}\to Z_\tau^{1/2}$ reads
\begin{equation}
   \left\langle\!\left\langle
   \phi(p_1e^\tau)\dotsb\phi(p_ne^\tau)
   \right\rangle\!\right\rangle_{S_\tau}^{K,k}
   =e^{-\tau n(D+2)/2}Z_\tau^{n/2}
   \left\langle\!\left\langle
   \phi(p_1)\dotsb\phi(p_n)
   \right\rangle\!\right\rangle_{S_{\tau=0}}^{K,k}.
\label{eq:(2.1)}
\end{equation}
The anomalous dimension in~Eq.~\eqref{eq:(1.6)} and the wave function
renormalization factor~$Z_\tau$ are related by
\begin{equation}
   \eta_\tau=\frac{\partial}{\partial\tau}\ln Z_\tau.
\label{eq:(2.2)}
\end{equation}
Here, the modified correlation functions are defined by~\cite{Sonoda:2015bla}
\begin{equation}
   \left\langle\!\left\langle
   \phi(p_1)\dotsb\phi(p_n)
   \right\rangle\!\right\rangle_S^{K,k}
   \equiv
   \prod_{i=1}^n\frac{1}{K(p_i)}
   \left\langle
   \exp\left[-\int_p\frac{k(p)}{p^2}
   \frac{1}{2}
   \frac{\delta^2}{\delta\phi(p)\delta\phi(-p)}\right]
   \phi(p_1)\dotsb\phi(p_n)
   \right\rangle_S,
\label{eq:(2.3)}
\end{equation}
where the ordinary correlation functions are denoted with single brackets:
\begin{equation}
   \left\langle
   \phi(p_1)\dotsb\phi(p_n)
   \right\rangle_S
   \equiv\int[d\phi]\,
   \phi(p_1)\dotsb\phi(p_n)
   \,e^{S[\phi]}.
\label{eq:(2.4)}
\end{equation}
In terms of ordinary correlation functions, Eq.~\eqref{eq:(2.1)} reads
\begin{align}
   &\left\langle
   \exp\left[-\int_p\frac{k(p)}{p^2}
   \frac{1}{2}
   \frac{\delta^2}{\delta\phi(p)\delta\phi(-p)}\right]
   \phi(p_1e^\tau)\dotsb\phi(p_ne^\tau)
   \right\rangle_{S_\tau}
\notag\\
   &=e^{-\tau n(D+2)/2}Z_\tau^{n/2}
   \prod_{i=1}^n\frac{K(p_ie^\tau)}{K(p_i)}
   \left\langle
   \exp\left[-\int_p\frac{k(p)}{p^2}
   \frac{1}{2}
   \frac{\delta^2}{\delta\phi(p)\delta\phi(-p)}\right]
   \phi(p_1)\dotsb\phi(p_n)
   \right\rangle_{S_{\tau=0}}.
\label{eq:(2.5)}
\end{align}

Now, let us choose the Gaussian
\begin{equation}
   K(p)=e^{-p^2}
\label{eq:(2.6)}
\end{equation}
as the cutoff function~$K$. We then have
\begin{align}
   &\left\langle
   \exp\left[-\int_p\frac{k(p)}{p^2}
   \frac{1}{2}
   \frac{\delta^2}{\delta\phi(p)\delta\phi(-p)}\right]
   \phi(p_1e^\tau)\dotsb\phi(p_ne^\tau)
   \right\rangle_{S_\tau}
\notag\\
   &=e^{-\tau n(D+2)/2}Z_\tau^{n/2}
   \left\langle
   \exp\left[-\int_p\frac{k(p)}{p^2}
   \frac{1}{2}
   \frac{\delta^2}{\delta\phi(p)\delta\phi(-p)}\right]
   \varphi(t,p_1)\dotsb\varphi(t,p_n)
   \right\rangle_{S_{\tau=0}},
\label{eq:(2.7)}
\end{align}
where
\begin{equation}
   \varphi(t,p)\equiv
   e^{-tp^2}\phi(p),\qquad
   t\equiv e^{2\tau}-1,
\label{eq:(2.8)}
\end{equation}
is the diffused scalar field in~Eq.~\eqref{eq:(1.4)} given in momentum space.
In terms of functional integrals, this reads
\begin{align}
   &\int[d\phi]\,
   \phi(p_1)\dotsb\phi(p_n)
   \exp\left[-\int_p\frac{k(p)}{p^2}
   \frac{1}{2}
   \frac{\delta^2}{\delta\phi(p)\delta\phi(-p)}\right]
   e^{S_\tau[\phi]}
\notag\\
   &=e^{-\tau n(D+2)/2}Z_\tau^{n/2}
\notag\\
   &\qquad{}
   \times\int[d\phi]\,
   \varphi(t,p_1e^{-\tau})
   \dotsb\varphi(t,p_ne^{-\tau})
   \exp\left[-\int_p\frac{k(p)}{p^2}
   \frac{1}{2}
   \frac{\delta^2}{\delta\phi(p)\delta\phi(-p)}\right]
   e^{S_{\tau=0}[\phi]}.
\label{eq:(2.9)}
\end{align}
Using field variables in coordinate space
\begin{equation}
   \phi(x)
   =\int_p\,e^{ipx}\,\phi(p),\qquad
   \varphi(t,x)
   =\int_p\,e^{ipx}\,\varphi(t,p),
\label{eq:(2.10)}
\end{equation}
we get $\delta/[\delta\phi(p)]=\int d^Dx\,e^{ipx}\,(\delta/[\delta\phi(x)])$ and
$\delta/[\delta\varphi(t,p)]=\int d^Dx\,e^{ipx}\,(\delta/[\delta\varphi(t,x)])$.
Hence, we can rewrite~Eq.~\eqref{eq:(2.9)} as
\begin{align}
   &\int[d\phi]\,
   \phi(x_1)\dotsb\phi(x_n)
   \exp\left[
   -\int d^Dx\int d^Dy\,\,\mathcal{D}(x-y)
   \frac{1}{2}
   \frac{\delta^2}{\delta\phi(x)\delta\phi(y)}\right]
   e^{S_\tau[\phi]}
\notag\\
   &=e^{\tau n(D-2)/2}Z_\tau^{n/2}
\notag\\
   &\qquad{}
   \times\int[d\phi]\,
   \varphi(t,x_1e^\tau)
   \dotsb\varphi(t,x_ne^\tau)
   \exp\left[
   -\int d^Dx\int d^Dy\,\,\mathcal{D}(x-y)
   \frac{1}{2}
   \frac{\delta^2}{\delta\phi(x)\delta\phi(y)}\right]
   e^{S_{\tau=0}[\phi]},
\label{eq:(2.11)}
\end{align}
where
\begin{equation}
   \mathcal{D}(x)\equiv\int_p\,e^{ipx}\,
   \frac{k(p)}{p^2}.
\label{eq:(2.12)}
\end{equation}
This leads to a representation of the Wilson action~$S_\tau[\phi]$,
\begin{align}
   e^{S_\tau[\phi]}
   &=\exp\left[
   \int d^Dx\int d^Dy\,\,\mathcal{D}(x-y)
   \frac{1}{2}
   \frac{\delta^2}{\delta\phi(x)\delta\phi(y)}\right]
\notag\\
   &\qquad{}
   \times\int[d\phi']\,
   \prod_{x'}\delta
   \left(\phi(x)
   -e^{\tau(D-2)/2}Z_\tau^{1/2}\varphi'(t,x'e^\tau)\right)
\notag\\
   &\qquad\qquad{}
   \times\exp\left[
   -\int d^Dx''\int d^Dy''\,\,\mathcal{D}(x''-y'')
   \frac{1}{2}
   \frac{\delta^2}{\delta\phi'(x'')\delta\phi'(y'')}\right]
   e^{S_{\tau=0}[\phi']}.
\label{eq:(2.13)}
\end{align}
Note that the field~$\varphi'(t,x'e^\tau)$ in the delta function results from
diffusion of the integration variable~$\phi'$ by the flow
equation~\eqref{eq:(1.4)}. It is easy to check Eq.~\eqref{eq:(2.13)} simply by
substituting it into~Eq.~\eqref{eq:(2.11)}. Written with the diffused field in
coordinate space, this representation admits straightforward generalization to
the other systems whose gradient flow equation may be non-linear in fields.
Yang--Mills theory is such an example.\footnote{We can also generalize this to
the $O(N)$ non-linear sigma model~\cite{Kikuchi:2014rla,Makino:2014sta,%
Makino:2014cxa,Aoki:2014dxa}.} Equation~\eqref{eq:(2.13)} is the basis of our
construction in the next section.

Before discussing generalization to Yang--Mills theory, let us verify that
Eq.~\eqref{eq:(2.13)} satisfies the ERG equation~\eqref{eq:(1.6)}. Recalling
$t=e^{2\tau}-1$ (Eq.~\eqref{eq:(2.8)}) and the flow equation~\eqref{eq:(1.4)},
we find
\begin{align}
   &\frac{\partial}{\partial\tau}\,e^{S_\tau[\phi]}
\notag\\
   &=\exp\left[
   \int d^Dx\int d^Dy\,\,\mathcal{D}(x-y)
   \frac{1}{2}
   \frac{\delta^2}{\delta\phi(x)\delta\phi(y)}\right]
\notag\\
   &\qquad{}
   \times\int[d\phi']\,
   \int d^Dx'\,
   \left[-\frac{D-2}{2}-\frac{\eta_\tau}{2}
   -2\Delta_{x'}
   -x_\mu'\frac{\partial}{\partial x_\mu'}\right]
   e^{\tau(D-2)/2}Z_\tau^{1/2}\varphi'(t,x'e^\tau)
\notag\\
   &\qquad\qquad{}
   \times\frac{\delta}{\delta\phi(x')}
   \prod_x\delta
   \left(\phi(x)
   -e^{\tau(D-2)/2}Z_\tau^{1/2}\varphi'(t,xe^\tau)\right)
\notag\\
   &\qquad\qquad\qquad{}
   \times\exp\left[
   -\int d^Dx\int d^Dy\,\,\mathcal{D}(x-y)
   \frac{1}{2}
   \frac{\delta^2}{\delta\phi'(x)\delta\phi'(y)}\right]
   e^{S_{\tau=0}[\phi']}
\notag\\
   &=\exp\left[
   \int d^Dx\int d^Dy\,\,\mathcal{D}(x-y)
   \frac{1}{2}
   \frac{\delta^2}{\delta\phi(x)\delta\phi(y)}\right]
\notag\\
   &\qquad{}
   \times
   \int d^Dx'\,\frac{\delta}{\delta\phi(x')}
   \left[-2\Delta_{x'}-\frac{D-2}{2}-\frac{\eta_\tau}{2}
   -x_\mu'\frac{\partial}{\partial x_\mu'}\right]
   \phi(x')
\notag\\
   &\qquad\qquad{}
   \times\int[d\phi']\,
   \prod_x\delta
   \left(\phi(x)
   -e^{\tau(D-2)/2}Z_\tau^{1/2}\varphi'(t,xe^\tau)\right)
\notag\\
   &\qquad\qquad\qquad{}
   \times\exp\left[
   -\int d^Dx\int d^Dy\,\,\mathcal{D}(x-y)
   \frac{1}{2}
   \frac{\delta^2}{\delta\phi'(x)\delta\phi'(y)}\right]
   e^{S_{\tau=0}[\phi']}
\notag\\
   &=\exp\left[
   \int d^Dx\int d^Dy\,\,\mathcal{D}(x-y)
   \frac{1}{2}
   \frac{\delta^2}{\delta\phi(x)\delta\phi(y)}\right]
\notag\\
   &\qquad{}
   \times
   \int d^Dx'\,
   \left[-2\Delta_{x'}-\frac{D-2}{2}-\frac{\eta_\tau}{2}
   -x_\mu'\frac{\partial}{\partial x_\mu'}\right]
   \phi(x')\cdot\frac{\delta}{\delta\phi(x')}
\notag\\
   &\qquad\qquad{}
   \times\int[d\phi']\,
   \prod_x\delta
   \left(\phi(x)
   -e^{\tau(D-2)/2}Z_\tau^{1/2}\varphi'(t,xe^\tau)\right)
\notag\\
   &\qquad\qquad\qquad{}
   \times\exp\left[
   -\int d^Dx\int d^Dy\,\,\mathcal{D}(x-y)
   \frac{1}{2}
   \frac{\delta^2}{\delta\phi'(x)\delta\phi'(y)}\right]
   e^{S_{\tau=0}[\phi']}.
\label{eq:(2.14)}
\end{align}
The first equality is obvious. In the second equality, we have made the
replacement, $e^{\tau(D-2)/2}Z_\tau^{1/2}\varphi'(t,x'e^\tau)\to\phi(x')$, which
is justified in front of the delta function. Then, we have interchanged
$\delta/[\delta\phi(x')]$ and~$\phi(x')$ neglecting an infinite constant
$\frac{\delta}{\delta\phi(x')}\phi(x')=\delta^{(D)}(x=0)$ because this
contributes only to the constant term in~$S_\tau[\phi]$. Finally, using the
relation
\begin{align}
   &\exp\left[
   \int d^Dx\int d^Dy\,\,\mathcal{D}(x-y)
   \frac{1}{2}
   \frac{\delta^2}{\delta\phi(x)\delta\phi(y)}\right]
   \phi(x')
\notag\\
   &=\left[\phi(x')
   +\int d^Dx\,\,\mathcal{D}(x-x')
   \frac{\delta}{\delta\phi(x)}\right]
   \exp\left[
   \int d^Dx\int d^Dy\,\,\mathcal{D}(x-y)
   \frac{1}{2}
   \frac{\delta^2}{\delta\phi(x)\delta\phi(y)}\right],
\label{eq:(2.15)}
\end{align}
we obtain an ERG equation
\begin{align}
   &\frac{\partial}{\partial\tau}\,e^{S_\tau[\phi]}
\notag\\
   &=\int d^Dx'\,
   \left(-2\Delta_{x'}-\frac{D-2}{2}-\frac{\eta_\tau}{2}
   -x_\mu'\frac{\partial}{\partial x_\mu'}\right)
   \left[\phi(x')
   +\int d^Dx\,\,\mathcal{D}(x-x')
   \frac{\delta}{\delta\phi(x)}\right]
\notag\\
   &\qquad\qquad\qquad\qquad\qquad{}
   \times\frac{\delta}{\delta\phi(x')}\,
   e^{S_\tau[\phi]}.
\label{eq:(2.16)}
\end{align}
Here, the derivative with respect to~$x'$ does not act on~$x'$
in~$\delta/[\delta\phi(x')]$. Switching back to momentum space, we get
\begin{align}
   \frac{\partial}{\partial\tau}\,
   e^{S_\tau[\phi]}
   &=\int_p\biggl\{
   \left[
   \left(2p^2+\frac{D+2}{2}-\frac{\eta_\tau}{2}\right)
   \phi(p)
   +p_\mu\frac{\partial}{\partial p_\mu}\phi(p)
   \right]
   \frac{\delta}{\delta\phi(p)}
\notag\\
   &\qquad\qquad{}
   +\frac{1}{p^2}\left[
   4p^2k(p)
   +2p^2\frac{dk(p)}{dp^2}
   -\eta_\tau k(p)
   \right]
   \frac{1}{2}
   \frac{\delta^2}{\delta\phi(p)\delta\phi(-p)}
   \biggr\}\,
   e^{S_\tau[\phi]}.
\label{eq:(2.17)}
\end{align}
Since $\Delta(p)$ in~Eq.~\eqref{eq:(1.8)} is given by~$2p^2e^{-p^2}$ for our
choice~\eqref{eq:(2.6)}, this equation coincides precisely with the ERG
equation in momentum space, Eq.~\eqref{eq:(1.6)}.

\section{Yang--Mills theory}
\label{sec:3}
\subsection{Wilson action}
\label{sec:3.1}
A natural generalization of~Eq.~\eqref{eq:(2.13)} to Yang--Mills theory is
given by
\begin{align}
   e^{S_\tau[A]}
   &=\exp\left[
   \int d^Dx\,
   \frac{1}{2}
   \frac{\delta^2}{\delta A_\mu^a(x)\delta A_\mu^a(x)}\right]
\notag\\
   &\qquad{}
   \times\int[dA']\,
   \prod_{x',\nu,b}\delta
   \left(A_\nu^b(x')-e^{\tau(D-2)/2}B_\nu^{\prime b}(t,x'e^\tau)\right)
\notag\\
   &\qquad\qquad{}
   \times
   \exp\left[
   -\int d^Dx''\,
   \frac{1}{2}
   \frac{\delta^2}
   {\delta A_\rho^{\prime c}(x'')\delta A_\rho^{\prime c}(x'')}\right]
   e^{S_{\tau=0}[A']},
\label{eq:(3.1)}
\end{align}
where, as in~Eq.~\eqref{eq:(2.8)}, we identify the flow time~$t$ and the scale
parameter~$\tau$ by
\begin{equation}
   t\equiv e^{2\tau}-1.
\label{eq:(3.2)}
\end{equation}
The field $B_\nu^{\prime b}(t,x'e^\tau)$ in the delta function is diffused from
the integration variable~$A'$ by the flow equation
\begin{equation}
   \partial_tB_\mu^a(t,x)=D_\nu G_{\nu\mu}^a(t,x)
   +\alpha_0 D_\mu\partial_\nu B_\nu^a(t,x),\qquad
   B_\mu^a(t=0,x)=A_\mu^a(x).
\label{eq:(3.3)}
\end{equation}
Note that we have added a ``gauge-fixing term'' with the
parameter~$\alpha_0$~\cite{Luscher:2010iy,Luscher:2011bx} to the original flow
equation~\eqref{eq:(1.1)}; this term suppresses the gauge degrees of freedom
along the diffusion and guarantees the finiteness of gauge non-invariant
correlation functions of the diffused gauge field in perturbation
theory~\cite{Luscher:2011bx}. This somewhat peculiar addition is due to our
tacit assumption of perturbation theory in this section. In fact, we exclude
this term in lattice gauge theory discussed in the next section. In
transcribing Eq.~\eqref{eq:(2.13)} to gauge theory, we have set~$Z_\tau=1$
because the diffused field does not receive wave function
renormalization~\cite{Luscher:2011bx}; we will see that this choice is
consistent with an effective presence of a cutoff in the Wilson action. We have
also adopted $k(p)=p^2$ which yields $\mathcal{D}(x)=\delta^{(D)}(x)$
in~Eq.~\eqref{eq:(2.12)}.

Under a change of the scale parameter~$\tau$, Eq.~\eqref{eq:(3.1)} preserves
the partition function:
\begin{align}
   \int[dA]\,e^{S_\tau[A]}
   &=\int[dA]\,\exp\left[
   -\int d^Dx\,
   \frac{1}{2}
   \frac{\delta^2}{\delta A_\mu^a(x)\delta A_\mu^a(x)}\right]
   e^{S_\tau[A]}
\notag\\
   &=\int[dA]\,\exp\left[
   -\int d^Dx\,
   \frac{1}{2}
   \frac{\delta^2}{\delta A_\mu^a(x)\delta A_\mu^a(x)}\right]
   e^{S_{\tau=0}[A]}
\notag\\
   &=\int[dA]\,e^{S_{\tau=0}[A]}.
\label{eq:(3.4)}
\end{align}
The first equality follows from the vanishing of a total derivative
$\int[dA]\,(\delta/\delta[A_\mu^a(x)])\,\mathcal{F}[A]=0$ for any
well-behaved functional~$\mathcal{F}[A]$; for the second equality, we have
used~Eq.~\eqref{eq:(3.1)}. The invariance of the partition function, expected
of a Wilson action, remains formal unless the functional integral in the most
right-hand side of~Eq.~\eqref{eq:(3.4)} is regularized. In perturbation theory,
at least, we can give a gauge-invariant meaning to the last integral by
dimensional regularization. With the lattice transcription
of~Eq.~\eqref{eq:(3.1)} in the next section, the invariance of the partition
function can be given a rigorous meaning.

Another important relation that follows immediately from~Eq.~\eqref{eq:(3.1)}
is
\begin{align}
   &\left\langle
   \exp\left[
   -\int d^Dx\,
   \frac{1}{2}
   \frac{\delta^2}{\delta A_\mu^a(x)\delta A_\mu^a(x)}\right]
   A_{\mu_1}^{a_1}(x_1)\dotsb A_{\mu_n}^{a_n}(x_n)
   \right\rangle_{S_\tau}
\notag\\
   &=e^{\tau n(D-2)/2}
   \left\langle
   \exp\left[
   -\int d^Dx\,
   \frac{1}{2}
   \frac{\delta^2}{\delta A_\mu^a(x)\delta A_\mu^a(x)}\right]
   B_{\mu_1}^{a_1}(t,x_1e^\tau)\dotsb B_{\mu_n}^{a_n}(t,x_ne^\tau)
   \right\rangle_{S_{\tau=0}}.
\label{eq:(3.5)}
\end{align}
This is analogous to~Eq.~\eqref{eq:(2.7)} in scalar field theory. As for the
right-hand side, note that the flow equation~\eqref{eq:(3.3)} can be written as
an integral equation~\cite{Luscher:2010iy,Luscher:2011bx}:
\begin{equation}
   B_\mu^a(t,x)
   =\int\mathrm{d}^Dy\left[
   K_t(x-y)_{\mu\nu}A_\nu^a(y)
   +\int_0^t\mathrm{d}s\,K_{t-s}(x-y)_{\mu\nu}R_\nu^a(s,y)
   \right],
\label{eq:(3.6)}
\end{equation}
where
\begin{equation}
   K_t(x)_{\mu\nu}\equiv\int_p\frac{\mathrm{e}^{ipx}}{p^2}
   \left[(\delta_{\mu\nu}p^2-p_\mu p_\nu)\mathrm{e}^{-tp^2}
   +p_\mu p_\nu \mathrm{e}^{-\alpha_0tp^2}\right]
\label{eq:(3.7)}
\end{equation}
is the integration kernel of a linear diffusion, and
\begin{equation}
   R_\mu^a\equiv
   f^{abc}
   \left[
   2B_\nu^b\partial_\nu B_\mu^c
   -B_\nu^b\partial_\mu B_\nu^c
   +(\alpha_0-1)B_\mu^b\partial_\nu B_\nu^c
   +f^{cde}B_\nu^bB_\nu^dB_\mu^e
   \right].
\label{eq:(3.8)}
\end{equation}
Using Eq.~\eqref{eq:(3.6)}, we can express $\delta B/\delta A$, necessary on
the right-hand side of~Eq.~\eqref{eq:(3.5)}, as a power series in~$B$. The
right-hand side of~Eq.~\eqref{eq:(3.5)} is then given by correlation functions
of the diffused field~$B$.

We now suppose that the ``bare'' action~$S_{\tau=0}[A]$ contains a gauge
coupling~$g_0$. Setting\footnote{Here,
$Z_g(\epsilon)=1-[g^2/(4\pi)^2](\beta_0/2\epsilon)+O(g^4)$
and~$\beta_0=(11/3)C_A$, where $C_A$ is the Casimir of the adjoint
representation, $f^{abc}f^{bcd}=C_A\delta^{ab}$.}
$g_0=\mu^\epsilon Z_g(\epsilon)g$, where $\mu$~is an arbitrary mass scale
and~$D=4-2\epsilon$, we take $\epsilon\to0$ for a continuum limit. By a general
theorem~\cite{Luscher:2011bx}, the right-hand of~Eq.~\eqref{eq:(3.5)} has a
finite limit. Hence, the correlation functions with respect to~$S_\tau[A]$ on
the left-hand side of~Eq.~\eqref{eq:(3.5)} are finite in the continuum limit.
This suggests that our definition of the Wilson action~\eqref{eq:(3.1)}
implements effectively an ultraviolet cutoff for the Wilson action.\footnote{In
a lattice transcription of~Eq.~\eqref{eq:(3.1)} in the next section, the
presence of an ultraviolet cutoff in the Wilson action is obvious.}

\subsection{Gauge invariance}
\label{sec:3.2}
We next show that $S_\tau[A]$ defined by~Eq.~\eqref{eq:(3.1)} is invariant
under any infinitesimal gauge transformation of the scaled gauge potential
\begin{equation}
   \widetilde{A_\mu^a}(x)\equiv\lambda A_\mu^a(x),
\label{eq:(3.9)}
\end{equation}
where
\begin{equation}
   \lambda\equiv e^{-\tau(D-4)/2}.
\label{eq:(3.10)}
\end{equation}
The $\tau$-dependent factor $\lambda$ acts like a coupling constant: An
infinitesimal gauge transformation on~$\widetilde{A}$ is
\begin{equation}
   \widetilde{A}_\mu^a(x)\longrightarrow
   \widetilde{A}_\mu^a(x)+\partial_\mu^x\omega^a(xe^\tau)
   +f^{abc}\widetilde{A}_\mu^b(x)\omega^c(xe^\tau),
\label{eq:(3.11)}
\end{equation}
but the corresponding gauge transformation on~$A$ is modified by~$\lambda$ as
\begin{equation}
   A_\mu^a(x)\longrightarrow
   A_\mu^a(x)+\lambda^{-1}\partial_\mu^x\omega^a(xe^\tau)
   +f^{abc}A_\mu^b(x)\omega^c(xe^\tau).
\label{eq:(3.12)}
\end{equation}
(See the Appendix for an alternative normalization of~$A$.)

To see the invariance of~$S_\tau[A]$, we first note that the first factor
in~Eq.~\eqref{eq:(3.1)}
\begin{equation}
   \exp\left[
   \int d^Dx\,
   \frac{1}{2}
   \frac{\delta^2}{\delta A_\mu^a(x)\delta A_\mu^a(x)}\right]
\label{eq:(3.13)}
\end{equation}
is invariant under the transformation~\eqref{eq:(3.12)} because the functional
derivative transforms in the adjoint representation
under~Eq.~\eqref{eq:(3.12)}:
\begin{equation}
   \frac{\delta}{\delta A_\mu^a(x)}
   \longrightarrow f^{abc}\frac{\delta}{\delta A_\mu^b(x)}\omega^c(xe^\tau).
\label{eq:(3.14)}
\end{equation}

We next examine the argument of the delta function in~Eq.~\eqref{eq:(3.1)}.
Under the transformation~\eqref{eq:(3.12)}, we find (we write $x'$ as~$x$ for
simplicity)
\begin{align}
   &A_\nu^b(x)-e^{\tau(D-2)/2}B_\nu^{\prime b}(t,xe^\tau)
\notag\\
   &\longrightarrow
   A_\nu^b(x)+\lambda^{-1}\partial_\nu^x\omega^b(xe^\tau)
   +f^{bcd}A_\nu^c(x)\omega^d(xe^\tau)
   -e^{\tau(D-2)/2}B_\nu^{\prime b}(t,xe^\tau)
\notag\\
   &=A_\nu^b(x)
   -e^{\tau(D-2)/2}\left[
   B_\nu^{\prime b}(t,xe^\tau)
   -e^{-\tau}\partial_\nu^x\omega^b(xe^\tau)
   -f^{bcd}e^{-\tau(D-2)/2}A_\nu^c(x)\omega^d(xe^\tau)
   \right]
\notag\\
   &=A_\nu^b(x)-e^{\tau(D-2)/2}
   \left[
   B_\nu^{\prime b}(t,xe^\tau)
   -\partial_\nu\omega^b(xe^\tau)
   -f^{bcd}B_\nu^{\prime c}(t,xe^\tau)\omega^d(xe^\tau)
   \right]
\notag\\
   &=A_\nu^b(x)-e^{\tau(D-2)/2}
   \left[
   B_\nu^{\prime b}(t,xe^\tau)
   -D_\nu'\omega^b(xe^\tau)
   \right].
\label{eq:(3.15)}
\end{align}
In the third line above, we can replace $e^{-\tau(D-2)/2}A_\nu^c(x)$
by~$B_\nu^{\prime c}(t,xe^\tau)$ since $\omega$ is infinitesimal, and the two are
equal when $\omega=0$. The last line implies that the gauge
transformation~\eqref{eq:(3.12)} on the external variable~$A$ induces a gauge
transformation on~$B_\nu^{\prime b}(t,xe^\tau)$ with the gauge
function~$-\omega^b(xe^\tau)$:
\begin{equation}
   B_\mu^{\prime a}(t,x)\longrightarrow  B_\mu^{\prime a}(t,x)-D_\mu'\omega^a(x).
\label{eq:(3.16)}
\end{equation}

In the functional integral~\eqref{eq:(3.1)}, the integration variable~$A'$ and
the diffused gauge field~$B'$ are related by the flow
equation~\eqref{eq:(3.3)}. We wish to show that there is a gauge transformation
on~$A'$ that gives the gauge transformed~$B'$, given by~Eq.~\eqref{eq:(3.16)},
as the solution of the diffusion equation~\eqref{eq:(3.3)}. To show this, let
us consider an infinitesimal gauge transformation on the diffused field~$B$
that \emph{depends\/} on the flow time~$s$ (we save $t$ for~$t=e^{2\tau}-1$):
\begin{equation}
   B_\mu^a(s,x)\longrightarrow B_\mu^a(s,x)-D_\mu\xi^a(s,x).
\label{eq:(3.17)}
\end{equation}
This changes the flow equation~\eqref{eq:(3.3)} to
\begin{equation}
   \partial_s B_\mu^a(s,x)
   =D_\nu G_{\nu\mu}^a(s,x)
   +\alpha_0 D_\mu\partial_\nu B_\nu^a(s,x)
   +D_\mu(\partial_s-\alpha_0D_\nu\partial_\nu)\xi^a(s,x).
\label{eq:(3.18)}
\end{equation}
If we choose $\xi$ as the solution to the linear diffusion equation,
\begin{equation}
   (\partial_s-\alpha_0D_\nu\partial_\nu)\xi^a(s,x)=0,\qquad
   \xi^a(s=t,x)=\omega^a(x),
\label{eq:(3.19)}
\end{equation}
Eq.~\eqref{eq:(3.18)} reduces to the original diffusion
equation~\eqref{eq:(3.3)} (with $s$ replacing~$t$). Note that we must solve
Eq.~\eqref{eq:(3.19)} \emph{backward\/} against the flow time; $\xi$~is
specified at~$s=t$ rather than the usual $s=0$. Thus, if we gauge-transform the
integration variable~$A'$ by
\begin{equation}
   A_\mu^{\prime a}(x)\longrightarrow
   A_\mu^{\prime a}(x)-D_\mu'\xi^a(s=0,x),
\label{eq:(3.20)}
\end{equation}
the diffusion equation~\eqref{eq:(3.3)} gives the gauge-transformed~$B'$ given
by~Eq.~\eqref{eq:(3.16)}.

We have shown that the gauge transformation~\eqref{eq:(3.12)} on the external
variable~$A$ induces the ordinary gauge transformation~\eqref{eq:(3.20)} on the
integration variable~$A'$. Now, the functional measure~$[dA']$
in~Eq.~\eqref{eq:(3.1)} can be and is defined to be gauge invariant (by
dimensional regularization, for example). The factor
\begin{equation}
   \exp\left[
   -\int d^Dx''\,
   \frac{1}{2}
   \frac{\delta^2}
   {\delta A_\rho^{\prime c}(x'')\delta A_\rho^{\prime c}(x'')}\right]
\label{eq:(3.21)}
\end{equation}
is invariant just as the factor~\eqref{eq:(3.13)} is. We thus conclude that, if
the original ``bare'' action~$S_{\tau=0}[A]$ in~Eq.~\eqref{eq:(3.1)} is
invariant under the gauge transformation, then the Wilson action~$S_\tau[A]$ is
invariant under the $\lambda$-dependent (hence $\tau$-dependent) gauge
transformation~\eqref{eq:(3.12)}. This is how our definition of the Wilson
action preserves manifest gauge invariance.\footnote{To compute the correlation
functions of elementary fields such as~Eq.~\eqref{eq:(3.5)} in perturbation
theory, we need to add a gauge-fixing term to~$S_{\tau=0}[A]$, which breaks the
gauge invariance. This breaking propagates to~$S_\tau[A]$. In lattice gauge
theory in the next section, however, such breaking of gauge invariance by gauge
fixing is unnecessary.}

\subsection{ERG equation}
\label{sec:3.3}
We now derive an ERG differential equation satisfied by the above Wilson
action~\eqref{eq:(3.1)}. By using Eqs.~\eqref{eq:(3.2)} and~\eqref{eq:(3.3)},
calculations analogous to~Eq.~\eqref{eq:(2.14)} yield
\begin{align}
   &\frac{\partial}{\partial\tau}\,e^{S_\tau[A]}
\notag\\
   &=\exp\left[
   \int d^Dx\,
   \frac{1}{2}
   \frac{\delta^2}{\delta A_\mu^a(x)\delta A_\mu^a(x)}\right]
\notag\\
   &\qquad{}
   \times\int d^Dx'\,
   \frac{\delta}{\delta\widetilde{A_\nu^b}(x')}
   \left[
   -2\widetilde{D_\rho F_{\rho\nu}^b}(x')
   -2\alpha_0\widetilde{D_\nu\partial_\rho A_\rho^b}(x')
   -\left(\frac{D-2}{2}+x_\rho'\partial_\rho'\right)
   \widetilde{A_\nu^b}(x')
   \right]
\notag\\
   &\qquad\qquad{}
   \times\int[dA']\,
   \prod_{x'',\rho,c}\delta
   \left(
   A_\rho^c(x'')-e^{\tau(D-2)/2}B_\rho^{\prime c}(t,x''e^\tau)
   \right)
\notag\\
   &\qquad\qquad\qquad{}
   \times\exp\left[
   -\int d^Dx'''\,
   \frac{1}{2}
   \frac{\delta^2}{\delta A_\lambda^{\prime d}(x''')
   \delta A_\lambda^{\prime d}(x''')}
   \right]
   e^{S_{\tau=0}[A']}
\notag\\
   &=\exp\left[
   \lambda^2
   \int d^Dx\,
   \frac{1}{2}
   \frac{\delta^2}
   {\delta\widetilde{A_\mu^a}(x)\delta\widetilde{A_\mu^a}(x)}\right]
\notag\\
   &\qquad{}
   \times\int d^Dx'\,
   \frac{\delta}{\delta\widetilde{A_\nu^b}(x')}
   \left[
   -2\widetilde{D_\rho F_{\rho\nu}^b}(x')
   -2\alpha_0\widetilde{D_\nu\partial_\rho A_\rho^b}(x')
   -\left(\frac{D-2}{2}+x_\rho'\partial_\rho'
   \right)\widetilde{A_\nu^b}(x')
   \right]
\notag\\
   &\qquad\qquad{}
   \times\exp\left[
   -\lambda^2
   \int d^Dx''\,
   \frac{1}{2}
   \frac{\delta^2}
   {\delta \widetilde{A_\sigma^c}(x'')\delta\widetilde{A_\sigma^c}(x'')}\right]
   e^{S_\tau[A]},
\label{eq:(3.22)}
\end{align}
where the gauge potential~$A_\mu^a(x)$ under the tilde
($\widetilde{\phantom{x}}$) is replaced by the rescaled potential,
Eq.~\eqref{eq:(3.9)}.

Using a relation analogous to~Eq.~\eqref{eq:(2.15)} (with~$\delta^{(D)}(x)$
replacing $\mathcal{D}(x)$):
\begin{equation}
   \exp\left[
   \lambda^2
   \int d^Dx\,\frac{1}{2}
   \frac{\delta^2}
   {\delta\widetilde{A_\mu^a}(x)\delta\widetilde{A_\mu^a}(x)}\right]
   \widetilde{A_\nu^b}(x')
   =\widehat{\widetilde{A_\nu^b}}(x')
   \exp\left[
   \lambda^2
   \int d^Dx\,\frac{1}{2}
   \frac{\delta^2}
   {\delta\widetilde{A_\mu^a}(x)\delta\widetilde{A_\mu^a}(x)}\right],
\label{eq:(3.23)}
\end{equation}
where we define the hat~($\widehat{\phantom{x}}$) by
\begin{equation}
   \widehat{\widetilde{A_\mu^a}}(x)\equiv
   \widetilde{A_\mu^a}(x)
   +\lambda^2\frac{\delta}{\delta\widetilde{A_\mu^a}(x)},
\label{eq:(3.24)}
\end{equation}
we can rewrite Eq.~\eqref{eq:(3.22)} compactly as
\begin{align}
   &\frac{\partial}{\partial\tau}\,e^{S_\tau[A]}
\notag\\
   &=\int d^Dx\,
   \frac{\delta}{\delta\widetilde{A_\mu^a}(x)}
   \left[
   -2\widehat{\widetilde{D_\nu F_{\nu\mu}^a}}(x)
   -2\alpha_0\widehat{\widetilde{D_\mu\partial_\nu A_\nu^b}}(x)
   -\left(\frac{D-2}{2}+x_\nu\partial_\nu\right)
   \widehat{\widetilde{A_\mu^a}}(x)
   \right]
   e^{S_\tau[A]}.
\label{eq:(3.25)}
\end{align}
Here, the gauge potential~$\widetilde{A_\mu^a}(x)$ is replaced by the
combination~\eqref{eq:(3.24)} if it appears under the hat. This is our ERG
equation for Yang--Mills theory.

Note that without the hat, Eq.~\eqref{eq:(3.25)} would involve only the first
order differentials of~$S_\tau$, and our ERG equation would be merely a change
of variables. It is the differential operator in the hat~\eqref{eq:(3.24)},
whose origin is the exponentiated second-order differentials
in~Eq.~\eqref{eq:(3.22)}, that introduces higher-order differentials
in~Eq.~\eqref{eq:(3.25)}.

Once the ERG equation~\eqref{eq:(3.25)} has been obtained, we may forget the
original construction~\eqref{eq:(3.1)} and the gradient flow behind it. Under
the ERG flow, the gauge invariance is preserved in the sense explained
in~Sect.~\ref{sec:3.2}.

For completeness, we give a little more explicit form of the ERG
equation~\eqref{eq:(3.25)}:
\begin{align}
   &\frac{\partial}{\partial\tau}e^{S_\tau[A]}
\notag\\
   &=\int d^Dx\,\frac{\delta}{\delta\widetilde{A_\mu^a}(x)}
\notag\\
   &\qquad{}
   \times
   \Biggl\{
   -2\widetilde{D_\nu}
   \left[
   \widetilde{F_{\nu\mu}^a}(x)
   +\lambda^2\widetilde{D_\nu}\frac{\delta}{\delta\widetilde{A_\mu^a}(x)}
   -\lambda^2\widetilde{D_\mu}\frac{\delta}{\delta\widetilde{A_\nu^a}(x)}
   +\lambda^4
   f^{abc}
   \frac{\delta}{\delta\widetilde{A_\nu^b}(x)}
   \frac{\delta}{\delta\widetilde{A_\mu^c}(x)}
   \right]
\notag\\
   &\qquad\qquad{}
   -2\lambda^2f^{abc}\frac{\delta}{\delta\widetilde{A_\nu^b}(x)}
   \Biggl[
   \widetilde{F_{\nu\mu}^c}(x)
   +\lambda^2\widetilde{D_\nu}\frac{\delta}{\delta\widetilde{A_\mu^c}(x)}
   -\lambda^2\widetilde{D_\mu}\frac{\delta}{\delta\widetilde{A_\nu^c}(x)}
   +\lambda^4
   f^{cde}
   \frac{\delta}{\delta\widetilde{A_\nu^d}(x)}
   \frac{\delta}{\delta\widetilde{A_\mu^e}(x)}
   \Biggr]
\notag\\
   &\qquad\qquad{}
   -2\alpha_0
   \Biggl[
   \widetilde{D_\mu\partial_\nu A_\nu^a}(x)
   +\lambda^2
   \partial_\mu\partial_\nu\frac{\delta}{\delta\widetilde{A_\nu^a}(x)}
   +\lambda^2f^{abc}\widetilde{A_\mu^b}(x)
   \partial_\nu
   \frac{\delta}{\delta\widetilde{A_\nu^c}(x)}
\notag\\
   &\qquad\qquad\qquad\qquad{}
   +\lambda^2f^{abc}\frac{\delta}{\delta\widetilde{A_\mu^b}(x)}
   \partial_\nu\widetilde{A_\nu^c}(x)
   +\lambda^4
   f^{abc}
   \frac{\delta}{\delta\widetilde{A_\mu^b}(x)}
   \partial_\nu
   \frac{\delta}{\delta\widetilde{A_\nu^c}(x)}
   \Biggr]
\notag\\
   &\qquad\qquad{}
   -\left(\frac{D-2}{2}+x_\nu\partial_\nu\right)
   \left[
   \widetilde{A_\mu^a}(x)+\lambda^2\frac{\delta}{\delta\widetilde{A_\mu^a}(x)}
   \right]
   \Biggr\}\,
   e^{S_\tau[A]}.
\label{eq:(3.26)}
\end{align}
In deriving this, we have interchanged the order
of~$\delta/[\delta\widetilde{A_\nu^b}(x)]$ and~$\widetilde{A_\mu^c}(x)$ in
the
combination~$f^{abc}(\delta/[\delta\widetilde{A_\nu^b}(x)])\widetilde{A_\mu^c}(x)$; this is justified because~$f^{abc}$ is anti-symmetric
in~$b\leftrightarrow c$.

To write a differential equation for~$S_\tau$, we multiply $e^{-S_\tau}$ from the
left of~Eq.~\eqref{eq:(3.26)} and write covariant derivatives explicitly to
obtain
\begin{align}
   &\frac{\partial}{\partial\tau}S_\tau[A]
\notag\\
   &=e^{-S_\tau[A]}\int d^Dx\,
   \frac{\delta}{\delta A_\mu^a(x)}
\notag\\
   &\qquad{}
   \times
   \biggl\{
   -2\partial_\nu
   \biggl[
   \partial_\nu A_\mu^a(x)-\partial_\mu A_\nu^a(x)
   +\partial_\nu\frac{\delta}{\delta A_\mu^a(x)}
   -\partial_\mu\frac{\delta}{\delta A_\nu^a(x)}
\notag\\
   &\qquad\qquad\qquad\qquad{}
   +\lambda f^{abc}A_\nu^b(x)A_\mu^c(x)
   +\lambda f^{abc}
   \left[
   A_\nu^b(x)\frac{\delta}{\delta A_\mu^c(x)}
   -A_\mu^b(x)\frac{\delta}{\delta A_\nu^c(x)}
   \right]
\notag\\
   &\qquad\qquad\qquad\qquad{}
   +\lambda
   f^{abc}
   \frac{\delta}{\delta A_\nu^b(x)}
   \frac{\delta}{\delta A_\mu^c(x)}
   \biggr]
\notag\\
   &\qquad\qquad{}
   -2\lambda f^{abc}
   \left[
   A_\nu^b(x)+\frac{\delta}{\delta A_\nu^b(x)}
   \right]
\notag\\
   &\qquad\qquad\qquad{}
   \times\biggl[
   \partial_\nu A_\mu^c(x)-\partial_\mu A_\nu^c(x)
   +\partial_\nu\frac{\delta}{\delta A_\mu^c(x)}
   -\partial_\mu\frac{\delta}{\delta A_\nu^c(x)}
\notag\\
   &\qquad\qquad\qquad\qquad{}
   +\lambda f^{cde}A_\nu^d(x)A_\mu^e(x)
   +\lambda f^{cde}
   \left[
   A_\nu^d(x)\frac{\delta}{\delta A_\mu^e(x)}
   -A_\mu^d(x)\frac{\delta}{\delta A_\nu^e(x)}
   \right]
\notag\\
   &\qquad\qquad\qquad\qquad{}
   +\lambda
   f^{cde}
   \frac{\delta}{\delta A_\nu^d(x)}
   \frac{\delta}{\delta A_\mu^e(x)}
   \biggr]
\notag\\
   &\qquad\qquad{}
   -2\alpha_0
   \Biggl[
   \partial_\mu\partial_\nu A_\nu^a(x)
   +\partial_\mu\partial_\nu\frac{\delta}{\delta A_\nu^a(x)}
   +\lambda f^{abc}A_\mu^b(x)\partial_\nu A_\nu^c(x)
   +\lambda f^{abc}A_\mu^b(x)
   \partial_\nu
   \frac{\delta}{\delta A_\nu^c(x)}
\notag\\
   &\qquad\qquad\qquad\qquad{}
   +\lambda f^{abc}\frac{\delta}{\delta A_\mu^b(x)}
   \partial_\nu A_\nu^c(x)
   +\lambda
   f^{abc}
   \frac{\delta}{\delta A_\mu^b(x)}
   \partial_\nu
   \frac{\delta}{\delta A_\nu^c(x)}
   \Biggr]
\notag\\
   &\qquad\qquad{}
   -\left(\frac{D-2}{2}+x_\nu\partial_\nu\right)
   \left[
   A_\mu^a(x)+\frac{\delta}{\delta A_\mu^a(x)}
   \right]
   \biggr\}\,
   e^{S_\tau[A]}.
\label{eq:(3.27)}
\end{align}
Differentiating $e^{S_\tau}$ further, we obtain a non-linear ERG equation that
involves up to quartic differentials of~$S_\tau$:
\begin{align}
   &\frac{\partial}{\partial\tau}S_\tau[A]
\notag\\
   &=\int d^Dx\,
   \left[
   \frac{\delta S_\tau}{\delta A_\mu^a(x)}
   +\frac{\delta}{\delta A_\mu^a(x)}
   \right]
\notag\\
   &\qquad{}
   \times
   \Biggl(
   -2\partial_\nu
   \biggl\{
   \partial_\nu A_\mu^a(x)-\partial_\mu A_\nu^a(x)
   +\partial_\nu\frac{\delta S_\tau}{\delta A_\mu^a(x)}
   -\partial_\mu\frac{\delta S_\tau}{\delta A_\nu^a(x)}
\notag\\
   &\qquad\qquad\qquad\qquad{}
   +\lambda f^{abc}A_\nu^b(x)A_\mu^c(x)
   +\lambda f^{abc}
   \left[
   A_\nu^b(x)\frac{\delta S_\tau}{\delta A_\mu^c(x)}
   -A_\mu^b(x)\frac{\delta S_\tau}{\delta A_\nu^c(x)}
   \right]
\notag\\
   &\qquad\qquad\qquad\qquad{}
   +\lambda
   f^{abc}
   \left[
   \frac{\delta^2S_\tau}{\delta A_\nu^b(x)\delta A_\mu^c(x)}
   +\frac{\delta S_\tau}{\delta A_\nu^b(x)}
   \frac{\delta S_\tau}{\delta A_\mu^c(x)}
   \right]
   \biggr\}
\notag\\
   &\qquad\qquad{}
   -2\lambda f^{abc}
   \left[
   A_\nu^b(x)
   +\frac{\delta S_\tau}{\delta A_\nu^b(x)}
   +\frac{\delta}{\delta A_\nu^b(x)}
   \right]
\notag\\
   &\qquad\qquad\qquad{}
   \times\biggl\{
   \partial_\nu A_\mu^c(x)-\partial_\mu A_\nu^c(x)
   +\partial_\nu\frac{\delta S_\tau}{\delta A_\mu^c(x)}
   -\partial_\mu\frac{\delta S_\tau}{\delta A_\nu^c(x)}
\notag\\
   &\qquad\qquad\qquad\qquad{}
   +\lambda f^{cde}A_\nu^d(x)A_\mu^e(x)
   +\lambda f^{cde}
   \left[
   A_\nu^d(x)\frac{\delta S_\tau}{\delta A_\mu^e(x)}
   -A_\mu^d(x)\frac{\delta S_\tau}{\delta A_\nu^e(x)}
   \right]
\notag\\
   &\qquad\qquad\qquad\qquad{}
   +\lambda
   f^{cde}
   \left[
   \frac{\delta^2S_\tau}{\delta A_\nu^d(x)\delta A_\mu^e(x)}
   +\frac{\delta S_\tau}{\delta A_\nu^d(x)}
   \frac{\delta S_\tau}{\delta A_\mu^e(x)}
   \right]
   \biggr\}
\notag\\
   &\qquad\qquad{}
   -2\alpha_0
   \Biggl\{
   \partial_\mu\partial_\nu A_\nu^a(x)
   +\partial_\mu\partial_\nu\frac{\delta S_\tau}{\delta A_\nu^a(x)}
   +\lambda f^{abc}A_\mu^b(x)\partial_\nu A_\nu^c(x)
   +\lambda f^{abc}A_\mu^b(x)
   \partial_\nu
   \frac{\delta S_\tau}{\delta A_\nu^c(x)}
\notag\\
   &\qquad\qquad\qquad\qquad{}
   +\lambda f^{abc}
   \left[
   \frac{\delta S_\tau}{\delta A_\mu^b(x)}
   +\frac{\delta}{\delta A_\mu^b(x)}
   \right]
   \partial_\nu A_\nu^c(x)
\notag\\
   &\qquad\qquad\qquad\qquad{}
   +\lambda
   f^{abc}
   \left[
   \frac{\delta S_\tau}{\delta A_\mu^b(x)}
   \partial_\nu
   \frac{\delta S_\tau}{\delta A_\nu^c(x)}
   +\frac{\delta}{\delta A_\mu^b(x)}
   \partial_\nu
   \frac{\delta S_\tau}{\delta A_\nu^c(x)}
   \right]
   \Biggr\}
\notag\\
   &\qquad\qquad{}
   -\left(\frac{D-2}{2}+x_\nu\partial_\nu\right)
   \left[
   A_\mu^a(x)+\frac{\delta S_\tau}{\delta A_\mu^a(x)}
   \right]
   \Biggr).
\label{eq:(3.28)}
\end{align}

\subsection{Approximate solution to~$O(\lambda^0)$}
\label{sec:3.4}
From Eq.~\eqref{eq:(3.28)}, we see that the parameter~$\lambda$, whose original
definition is~Eq.~\eqref{eq:(3.10)}, provides a convenient expansion parameter
which organizes terms in the ERG equation. We expand the Wilson action in
powers of~$\lambda$ as
\begin{equation}
   S_\tau[A]\equiv
   \sum_{n=2}^\infty\lambda^{n-2}\frac{1}{n!}\int d^Dx_1\,\dotsb\int d^Dx_n\,
   w_{n,\mu_1\dotsb \mu_n}^{a_1\dotsb a_n}(x_1,\dotsc,x_n)
   A_{\mu_1}^{a_1}(x_1)\dotsb A_{\mu_n}^{a_n}(x_n),
\label{eq:(3.29)}
\end{equation}
where~$w_n=O(\lambda^0)$. By substituting this into the right-hand side
of~Eq.~\eqref{eq:(3.28)}, we obtain terms of the form
\begin{equation}
   \sum_{n=2}^\infty\lambda^{n-2}\frac{1}{n!}\int d^Dx_1\,\dotsb\int d^Dx_n\,
   W_{n,\mu_1\dotsb \mu_n}^{a_1\dotsb a_n}(x_1,\dotsc,x_n)
   A_{\mu_1}^{a_1}(x_1)\dotsb A_{\mu_n}^{a_n}(x_n).
\label{eq:(3.30)}
\end{equation}
Therefore, the expansion of the Wilson action in the form~\eqref{eq:(3.29)} is
consistent with the ERG equation~\eqref{eq:(3.28)}.

In this paper, we study only the lowest-order $O(\lambda^0)$ terms in some
detail,\footnote{This is the only term for the abelian gauge theory.}
postponing the higher-order calculations for future studies. We thus set
\begin{equation}
   S_\tau[A]
   =\frac{1}{2}\int d^Dx\int d^Dy\,
   w_{2,\mu\nu}^{ab}(x,y)A_\mu^a(x)A_\nu^b(y).
\label{eq:(3.31)}
\end{equation}
Equation~\eqref{eq:(3.28)} then gives
\begin{align}
   &\frac{\partial}{\partial\tau}\frac{1}{2}w_{2,\mu\nu}^{ab}(x,y)
\notag\\
   &=-2\partial_\rho\partial_\rho w_{2,\mu\nu}^{ab}(x,y)
   +(1-\alpha_0)\left[\partial_\mu\partial_\rho w_{2,\rho\nu}^{ab}(x,y)
   +\partial_\nu\partial_\rho w_{2,\mu\rho}^{ab}(x,y)\right]
\notag\\
   &\qquad{}
   +\left[\frac{D+2}{2}
   +\frac{1}{2}(x-y)_\rho\partial_\rho\right]
   w_{2,\mu\nu}^{ab}(x,y)
\notag\\
   &\qquad{}
   +\int d^Dz\,
   w_{2,\mu\rho}^{ac}(x,z)
   \left[
   \delta_{\rho\sigma}
   (-2\partial_\lambda^z\partial_\lambda^z+1)
   +2(1-\alpha_0)\partial_\rho^z\partial_\sigma^z
   \right]
   w_{2,\sigma\nu}^{cb}(z,y).
\label{eq:(3.32)}
\end{align}
In deriving this, we have neglected $\delta^{(D)}(x=0)$ assuming dimensional
regularization. Imposing the translational and rotational invariance and
global gauge invariance, we can write
\begin{equation}
   w_{2,\mu\nu}^{ab}(x,y)=\delta^{ab}\int_pe^{ip(x-y)}
   \left[T(p)(p^2\delta_{\mu\nu}-p_\mu p_\nu)+L(p)p_\mu p_\nu\right],
\label{eq:(3.33)}
\end{equation}
where $T(p)$ and~$L(p)$ are functions of~$p^2$. Equation~\eqref{eq:(3.32)}
then gives
\begin{align}
   \frac{1}{2}\frac{\partial}{\partial\tau}T
   &=-p^2\frac{\partial}{\partial p^2}T
   +p^2(2p^2+1)T^2
   +2p^2T,
\notag\\
   \frac{1}{2}\frac{\partial}{\partial\tau}L
   &=-p^2\frac{\partial}{\partial p^2}L
   +p^2(2\alpha_0p^2+1)L^2
   +2\alpha_0p^2L.
\label{eq:(3.34)}
\end{align}
The general solution is given by
\begin{equation}
   T(\tau,p)=-\frac{1}{C(pe^{-\tau})e^{-2p^2}+p^2},\qquad
   L(\tau,p)=-\frac{1}{D(pe^{-\tau})e^{-2\alpha_0p^2}+p^2},
\label{eq:(3.35)}
\end{equation}
where $C(p)$ and~$D(p)$ are arbitrary functions of~$p^2$. Locality demands that
$C(p)$ and~$D(p)$ can be expanded in powers of~$p^2$ at~$p=0$:
\begin{equation}
  C(p)=C_0+C_1p^2+\frac{1}{2}C_2(p^2)^2+\dotsb,\qquad
  D(p)=D_0+D_1p^2+\frac{1}{2}D_2(p^2)^2+\dotsb.
\label{eq:(3.36)}
\end{equation}
Unitary demands $C_0>0$ and~$D_0>0$.

As~$\tau\to+\infty$, the action $S_\tau[A]$ approaches an infrared fixed
point~$S^*[A]$, corresponding to constants $C_0$ and~$D_0$:
\begin{equation}
   T^*(p)=-\frac{1}{C_0e^{-2p^2}+p^2},\qquad
   L^*(p)=-\frac{1}{D_0e^{-2\alpha_0p^2}+p^2}.
\label{eq:(3.37)}
\end{equation}
Since $C_0>0$ and~$D_0>0$ are arbitrary, their variations give marginal
operators:
\begin{equation}
   \delta T(p)=\frac{\delta C_0e^{-2p^2}}{(C_0e^{-2p^2}+p^2)^2},\qquad
   \delta L(p)=\frac{\delta D_0e^{-2\alpha_0p^2}}{(D_0e^{-2\alpha_0p^2}+p^2)^2}.
\label{eq:(3.38)}
\end{equation}
It can be seen that these correspond to the change of normalization of the
gauge field~$A$ (see the Appendix).\footnote{$\delta D_0$ corresponds to an
infinitesimal change of the gauge-fixing parameter.} Infinitesimal $C_n$
and~$D_n$, on the other hand, give
\begin{align}
   \delta T(\tau,p)
   &\equiv T(\tau,p)-T^*(p)
   \simeq
   \frac{C_n(p^2e^{-2\tau})^ne^{-2p^2}}{(C_0e^{-2p^2}+p^2)^2},
\notag\\
   \delta L(\tau,p)
   &\equiv L(\tau,p)-L^*(p)
   \simeq
   \frac{D_n(p^2e^{-2\tau})^ne^{-2\alpha_0p^2}}{(D_0e^{-2\alpha_0p^2}+p^2)^2},
\label{eq:(3.39)}
\end{align}
where $n=1$, $2$, \dots, which correspond to irrelevant operators at the fixed
point.

If we make the particular choice of $C_0=1$ and~$D_0=\infty$
in~Eq.~\eqref{eq:(3.36)}, the fixed-point action becomes transverse:
\begin{equation}
   S_\tau^*[A]
   =-\frac{1}{2}\int d^Dx\int d^Dy\,
   \int_pe^{ip(x-y)}
   \frac{1}{e^{-2p^2}+p^2}
   (p^2\delta_{\mu\nu}-p_\mu p_\nu)
   A_\mu^a(x)A_\nu^a(y),
\label{eq:(3.40)}
\end{equation}
and the marginal operator at the fixed point is given by
\begin{equation}
   \mathcal{O}_0
   =\int d^Dx\int d^Dy\,
   \int_pe^{ip(x-y)}
   \frac{e^{-2p^2}}{(e^{-2p^2}+p^2)^2}
   (p^2\delta_{\mu\nu}-p_\mu p_\nu)
   A_\mu^a(x)A_\nu^a(y).
\label{eq:(3.41)}
\end{equation}

It is important to pursue the above analysis to higher orders in~$\lambda$ to
see how the ordinary beta function arises in our formalism.

\section{Lattice gauge theory}
\label{sec:4}
In the previous section, we have constructed a gauge-invariant Wilson action
and its associated ERG equation for a generic Yang--Mills theory in
continuum~$\mathbb{R}^4$. We now tailor the construction for lattice gauge
theory.\footnote{Many versions of the renormalization group transformation have
been proposed for lattice gauge theory. We cite
Refs.~\cite{Wilson:1979wp,Iwasaki:2011np} as the pioneering works. Some
of the more recent works are~Refs.~\cite{deForcrand:1999bi,Ejiri:2003sw}.} For
simplicity, we consider an infinite volume lattice~$\mathbb{Z}^4$. The discrete
coordinates on~$\mathbb{Z}^4$ render our ERG transformation discrete. This
discreteness is introduced through ``block-spins.''  Let us pick a fixed
``block-spin'' factor~$b$ from one of the integers $2$, $3$, \dots. We then
define a ``block-spin'' link variable by
\begin{equation}
   \mathcal{U}(x,\mu)
   \equiv U(x,\mu)U(x+\Hat{\mu},\mu)\dotsb U(x+(b-1)\Hat{\mu},\mu),\qquad
   x\in b\mathbb{Z}^4,
\label{eq:(4.1)}
\end{equation}
where $U(x,\mu)$ is a conventional link variable on the~$\mathbb{Z}^4$ lattice;
here, $\Hat{\mu}$ denotes the unit vector in the $\mu$~direction. This
$\mathcal{U}(x,\mu)$ is regarded as a link variable on the coarse
lattice~$b\mathbb{Z}^4$ scaled by the factor~$b$.

We then divide the range of the scale factor~$\tau$, originally continuous
in~$0\leq\tau<\infty$, into the contiguous intervals
\begin{equation}
   n\Delta\tau<\tau\leq(n+1)\Delta\tau,\qquad n=0,1,2,\dotsc,
\label{eq:(4.2)}
\end{equation}
where
\begin{equation}
   \Delta\tau\equiv\ln b.
\label{eq:(4.3)}
\end{equation}
The $n$th interval corresponds to the scaling of~$x$ by a factor between $b^n$
and~$b^{n+1}$. Multiplying a lattice coordinate~$x\in\mathbb{Z}^4$
by~$e^{\Delta\tau}=b$ gives the coordinate~$bx$ on the coarse
lattice~$b\mathbb{Z}^4$.

Now, we consider a continuous change of the Wilson action within one of the
intervals in~Eq.~\eqref{eq:(4.2)}. A natural extension of~Eq.~\eqref{eq:(3.1)}
for the interval $\tau=(n\Delta\tau,(n+1)\Delta\tau]$ would be the discrete
transformation from $S_n$ to~$S_{n+1}$, given by\footnote{Note that the
formula~\eqref{eq:(3.1)} can be used to relate the Wilson actions between two
non-zero $\tau$ values.}
\begin{align}
   e^{S_{n+1}[U]}
   &=\exp\left(
   \sum_{x,\mu,a}\frac{1}{2}\partial_{x,\mu}^a\partial_{x,\mu}^a
   \right)
   \int[dU']\,
   \prod_{x',\nu}\delta\left(U(x',\nu)-W_{\Delta\tau}'(bx',\nu)\right)
\notag\\
   &\qquad{}
   \times\exp\left(
   -\sum_{x'',\rho,b}\frac{1}{2}\partial_{x'',\rho}^b\partial_{x'',\rho}^b
   \right)
   e^{S_n[U']}.
\label{eq:(4.4)}
\end{align}
This needs a fair amount of explanation, which we give below.

First, $\partial_{x,\mu}^a$ is a link differential operator defined by (see also
Appendix~A of~Ref.~\cite{Luscher:2010iy})
\begin{equation}
   \partial_{x,\mu}^a\mathcal{F}[U]\equiv
   \left.\frac{d}{ds}\mathcal{F}[e^{sX}U]\right|_{s=0},\qquad
   X(y,\nu)=\begin{cases}
   T^a&\text{if $(y,\nu)=(x,\mu)$},\\
   0&\text{otherwise},\\
   \end{cases}
\label{eq:(4.5)}
\end{equation}
where $T^a$ denotes a (anti-hermitian) generator of the gauge group. The
exponentiated link differential operator in~Eq.~\eqref{eq:(4.4)} is an
analogue of the exponentiated functional differential operator
in~Eq.~\eqref{eq:(3.1)}.

Secondly, $W_\tau'(bx',\nu)$ in~Eq.~\eqref{eq:(4.4)} is the
solution of the lattice flow equation~\cite{Luscher:2009eq,Luscher:2010iy} on
the coarse lattice~$x\in b\mathbb{Z}^4$:
\begin{equation}
   \frac{\partial}{\partial\tau}W_\tau'(x,\mu)
   =-2\partial_{x,\mu}S_w[W_\tau']\cdot W_\tau'(x,\mu),
\label{eq:(4.6)}
\end{equation}
where $\partial_{x,\mu}\equiv T^a\partial_{x,\mu}^a$. The initial value
at~$\tau=0$ is given by the ``block-spin'' link
variable~\eqref{eq:(4.1)} constructed from the integration variable~$U'$
defined on~$\mathbb{Z}^4$:
\begin{equation}
   W_{\tau=0}'(x,\mu)
   =\mathcal{U}'(x,\mu)
   \equiv
   U'(x,\mu)U'(x+\Hat{\mu},\mu)\dotsb U'(x+(b-1)\Hat{\mu},\mu),\quad
   x\in b\mathbb{Z}^4.
\label{eq:(4.7)}
\end{equation}
It is the value of $W_\tau$ at~$\tau=\Delta\tau$ that appears in the delta
function. A possible choice of~$S_w[W]$ is the plaquette action,
\begin{equation}
   S_w[W]\equiv\sum_p\Re\tr[1-W(p)],
\label{eq:(4.8)}
\end{equation}
where the sum runs over the plaquettes~$p$ belonging to the coarse
lattice~$b\mathbb{Z}^4$, and $W(p)$~is the product of the ``block-spin'' link
variables around~$p$. Note that the lattice flow equation~\eqref{eq:(4.6)} is
written in terms of the scale factor~$\tau$ rather than the flow
time~$t=b^{2n}e^{2\tau}-1$. We have used $\partial/\partial t=%
b^{-2n}e^{-2\tau}(\partial/2\partial\tau)$ and absorbed the
factor~$b^{2n}e^{2\tau}$ into the right-hand side; this prescription is natural
because we have rescaled the lattice coordinates by the factor~$b^{2n}e^{2\tau}$
compared with~$n=0$. Thanks to this prescription, the ERG
transformation~\eqref{eq:(4.4)}~from $S_n$ to $S_{n+1}$ does not depend on~$n$
explicitly.

We obtain the lattice Wilson action~$S_{n+1}[U]$ by successive applications
of~Eq.~\eqref{eq:(4.4)} on the ``bare'' action~$S_0[U]$. The preservation of
the partition function and the gauge invariance, both demonstrated
in~Sect.~\ref{sec:3} on the basis of perturbation theory, now hold true
non-perturbatively, as we explain below.

First, we consider the partition function. If $[dU]$ is the group-invariant
Haar measure such that $[d(e^\eta U)]=[dU]$ for infinitesimal Lie algebra
elements~$\eta_\mu(x)$, we find, for any functional~$\mathcal{F}[U]$,
\begin{align}
   \int[dU]\,\mathcal{F}[U]
   &=\int[d(e^\eta U)]\,\mathcal{F}[e^\eta U]
\notag\\
   &=\int[dU]\,\mathcal{F}[e^\eta U]
\notag\\
   &=\int[dU]
   \left[\mathcal{F}[U]+\sum_x\eta_\mu^a(x)\partial_{x,\mu}^a\mathcal{F}[U]
   \right].
\label{eq:(4.9)}
\end{align}
This implies $\int[dU]\,\partial_{x,\mu}^a\mathcal{F}[U]=0$. Using this
identity for~Eq.~\eqref{eq:(4.4)}, we obtain
\begin{equation}
   \int[dU]\,e^{S_{n+1}[U]}=\int[dU]\,e^{S_n[U]}.
\label{eq:(4.10)}
\end{equation}
Hence, the partition function is preserved just as in~Eq.~\eqref{eq:(3.4)}.

As for the gauge invariance, we first note that a gauge transformation is
given by
\begin{equation}
   U(x,\mu)\longrightarrow
   U^g(x,\mu)\equiv g(x)U(x,\mu)g(x+\Hat{\mu})^{-1},\qquad
   g(x)\equiv e^{\omega(x)}.
\label{eq:(4.11)}
\end{equation}
If $\omega$ is infinitesimal, the link differential operator transforms in the
adjoint representation,
\begin{equation}
   \left(\partial_{x,\mu}^a\mathcal{F}[U]\right)_{U\to U^g}
   =\partial_{x,\mu}^a\mathcal{F}[U^g]
   +f^{abc}\omega^b(x)\partial_{x,\mu}^c\mathcal{F}[U],
\label{eq:(4.12)}
\end{equation}
where the link differential operator acts on~$U^g$ on the left-hand side, but
it acts on~$U$ of~$U^g$ on the right. This shows that
$(\partial_{x,\mu}^a\partial_{x,\mu}^a\mathcal{F}[U])_{U\to U^g}
=\partial_{x,\mu}^a\partial_{x,\mu}^a\mathcal{F}[U^g]$, and
in~Eq.~\eqref{eq:(4.4)} the gauge transformation on~$U$ and the first
exponentiated link differential operator commute.

The gauge transformation~\eqref{eq:(4.11)} acts on the delta function
in~Eq.~\eqref{eq:(4.4)} as (we set $x'\to x$ for simplicity)
\begin{align}
   &\delta\left(U(x,\nu)-W_{\Delta\tau}'(bx,\nu)\right)
\notag\\
   &\longrightarrow
   \delta\left(
   g(x)U(x,\nu)g(x+\Hat{\nu})^{-1}-W_{\Delta\tau}'(bx,\nu)
   \right)
\notag\\
   &=\delta\left(
   U(x,\nu)-g(x)^{-1}W_{\Delta\tau}'(bx,\nu)g(x+\Hat{\nu})
   \right).
\label{eq:(4.13)}
\end{align}
This shows that the gauge transformation~\eqref{eq:(4.11)} on~$U$ induces an
inverse gauge transformation~$W^{g^{-1}}_{\Delta \tau}$ on~$W_{\Delta\tau}'$ defined
on the coarse lattice~$b\mathbb{Z}^4$. Now, if $W_\tau'$ is the solution of the
lattice flow equation~\eqref{eq:(4.6)} with the initial
condition~$\mathcal{U}'$, given by Eq.~\eqref{eq:(4.7)}, then
$W_\tau^{\prime g^{-1}}$ is the solution with the initial
condition~$\mathcal{U}^{\prime g^{-1}}$ as long as $g$~does not depend on~$\tau$;
this follows from the property~\eqref{eq:(4.12)}. Hence, the gauge
transformation $g$ on~$U$ induces the inverse gauge transformation $g^{-1}$ on
the initial condition~$\mathcal{U}'$. To obtain this transformation
on~$b\mathbb{Z}^4$, we can introduce the following gauge transformation
on~$\mathbb{Z}^4$:
\begin{equation}
   U'(x,\mu)
   \longrightarrow
   h(x)^{-1}U'(x,\mu)h(x+\Hat{\mu}),\qquad
   h(x)=
   \begin{cases}
   g(y)&\text{if $x=by$ for~$y\in\mathbb{Z}^4$},\\
   1&\text{otherwise}.\\
   \end{cases}
\label{eq:(4.14)}
\end{equation}
This gauge transformation commutes with the second exponentiated link
differential operator in~Eq.~\eqref{eq:(4.4)} and, as long as $S_n[U']$ is
gauge invariant, the resulting Wilson action~$S_{n+1}[U]$ is also gauge
invariant. This completes our argument for the gauge invariance of the lattice
ERG transformation.

The structure of our Wilson action defined recursively by~Eq.~\eqref{eq:(4.4)}
resembles the ``lattice effective action'' that has been advocated and studied
in~Refs.~\cite{Kagimura:2015via,Yamamura:2015kva}. Our definition is different
in two crucial aspects, however: Eq.~\eqref{eq:(4.4)} has exponentiated link
differential operators, and the lattice points are rescaled in each step of
the ERG transformation. As we have emphasized in the previous section, these
two are essential ingredients for obtaining an ERG differential equation that
is non-linear in the Wilson action and entails scale transformation of space.

Finally, let us derive an ERG differential equation in lattice gauge theory
that follows from the definition~\eqref{eq:(4.4)} of the Wilson action. For
this, we define $S_{n+1}(\tau)[U]$ by
\begin{align}
   e^{S_{n+1}(\tau)[U]}
   &\equiv
   \exp\left(
   \sum_{x,\mu,a}\frac{1}{2}\partial_{x,\mu}^a\partial_{x,\mu}^a
   \right)
   \int[dU']\,
   \prod_{x',\nu}\delta\left(U(x',\nu)-W_\tau'(bx',\nu)\right)
\notag\\
   &\qquad{}
   \times\exp\left(
   -\sum_{x'',\rho,b}\frac{1}{2}\partial_{x'',\rho}^b\partial_{x'',\rho}^b
   \right)
   e^{S_n[U']}.
\label{eq:(4.15)}
\end{align}
We have introduced a diffusion factor~$\tau$ so that
\begin{equation}
   S_{n+1}(\Delta\tau)[U]=S_{n+1}[U].
\label{eq:(4.16)}
\end{equation}
As~$\tau\to0+$, $S_{n+1}(\tau)$ reduces essentially to~$S_n$, written for the
block-spin link variables~$\mathcal{U}$ defined by~Eq.~\eqref{eq:(4.7)}:
\begin{align}
   e^{S_{n+1}(\tau\to0+)[U]}
   &\equiv
   \exp\left(
   \sum_{x,\mu,a}\frac{1}{2}\partial_{x,\mu}^a\partial_{x,\mu}^a
   \right)
   \int[dU']\,
   \prod_{x',\nu}\delta\left(U(x',\nu)-\mathcal{U}'(bx',\nu)\right)
\notag\\
   &\qquad{}
   \times\exp\left(
   -\sum_{x'',\rho,b}\frac{1}{2}\partial_{x'',\rho}^b\partial_{x'',\rho}^b
   \right)
   e^{S_n[U']}.
\label{eq:(4.17)}
\end{align}
The dependence of~$S_{n+1}(\tau)$ on the diffusion factor~$\tau$ is given by
the differential equation,
\begin{align}
   &\frac{\partial}{\partial\tau}e^{S_{n+1}(\tau)[U]}
\notag\\
   &=\exp\left(
   \sum_{x,\mu,a}\frac{1}{2}\partial_{x,\mu}^a\partial_{x,\mu}^a
   \right)
   \int[dU']\,
   \sum_{y,\sigma,c}(-2)\partial_{y,\sigma}^cS_w[W_\tau']\cdot
   \partial_{y,\sigma}^{\prime c}
   \prod_{x',\nu}\delta\left(U(x',\nu)-W_\tau'(bx',\nu)\right)
\notag\\
   &\qquad\qquad{}
   \times\exp\left(
   -\sum_{x'',\rho,b}\frac{1}{2}\partial_{x'',\rho}^b\partial_{x'',\rho}^b
   \right)
   e^{S_n[U']}
\notag\\
   &=2\exp\left(
   \sum_{x,\mu,a}\frac{1}{2}\partial_{x,\mu}^a\partial_{x,\mu}^a
   \right)
   \sum_{y,\sigma,c}\partial_{y,\sigma}^c
   \left(
   \partial_{y,\sigma}^cS_w[U]\right)
   \int[dU']\,
   \prod_{x',\nu}\delta\left(U(x',\nu)-W_\tau'(bx',\nu)\right)
\notag\\
   &\qquad\qquad{}
   \times\exp\left(
   -\sum_{x'',\rho,b}\frac{1}{2}\partial_{x'',\rho}^b\partial_{x'',\rho}^b
   \right)
   e^{S_n[U']}.
\label{eq:(4.18)}
\end{align}
For the first equality above, we have used the lattice flow
equation~\eqref{eq:(4.6)} in evaluating $\frac{\partial}{\partial\tau}
\mathcal{F}[W_\tau']=\sum_{y,\sigma,c}[(\partial/\partial\tau)
W_\tau'(y,\sigma)\cdot W_\tau'(y,\sigma)]^c\,
\partial_{y,\sigma}^{\prime c}\mathcal{F}[W_\tau']$, which follows from the
definition of the link differential operator~\eqref{eq:(4.5)}. It is
understood that the operator~$\partial_{y,\sigma}^{\prime c}$ acts on~$W_\tau'$.
For the second equality, we have rewritten $\partial_{y,\sigma}^{\prime c}$ as
the derivative on~$U$, $\partial_{y,\sigma}^{\prime c}\to-\partial_{y,\sigma}^c$;
this identity holds because the link differential operator acts on the delta
function as $d/ds\,\delta(U(x',\nu)-e^{sT^a}W_\tau'(bx',\nu))
=d/ds\,\delta(e^{-sT^a}U(x',\nu)-W_\tau'(bx',\nu))$. This link differential
operator on~$U$ can be put outside to act on the integral over~$U'$. Then, we
can replace $\partial_{y,\sigma}^cS_w[W_\tau']$ by~$\partial_{y,\sigma}^cS_w[U]$
thanks to the delta function. Therefore, from~Eq.~\eqref{eq:(4.15)}, we get an
ERG differential equation
\begin{align}
   \frac{\partial}{\partial\tau}e^{S_{n+1}(\tau)[U]}
   &=\exp\left(
   \sum_{x,\mu,a}\frac{1}{2}\partial_{x,\mu}^a\partial_{x,\mu}^a
   \right)
   \sum_{x',\nu,b}\partial_{x',\nu}^b
   \left[
   \partial_{x',\nu}^bS_w[U]
   \right]
\notag\\
   &\qquad{}
   \times\exp\left(
   -\sum_{x'',\rho,c}\frac{1}{2}\partial_{x'',\rho}^c\partial_{x'',\rho}^c
   \right)
   e^{S_{n+1}(\tau)[U]}.
\label{eq:(4.19)}
\end{align}
By integrating this from~$\tau=0+$ to~$\tau=\Delta\tau$, we restore the finite
change of the Wilson action in~Eq.~\eqref{eq:(4.4)}.

Thus, our ERG transformation in lattice gauge theory consists of the rescaling
of lattice points by~Eq.~\eqref{eq:(4.17)} and the diffusion from~$\tau=0+$
to~$\tau=\Delta\tau$ by~Eq.~\eqref{eq:(4.19)} (see~Eq.~\eqref{eq:(4.16)}). As
we have shown, this transformation preserves the partition function and
manifest gauge-invariance of the Wilson action. It is important to note that
neither Eq.~\eqref{eq:(4.17)} nor~Eq.~\eqref{eq:(4.19)} depends explicitly
on~$n$. This implies a possibility of finding a fixed point solution,
$S_{n+1}=S_n$. The technique in~Ref.~\cite{Luscher:2009eq} appears helpful to
the study of such questions.

\section{Conclusion}
\label{sec:5}
Imitating the structure of the Wilson action in scalar field theory, expressed
by the field diffused by the flow equation, we have constructed a manifestly
gauge-invariant Wilson action and its associated ERG differential equation in
Yang--Mills theory. The construction, extended to lattice gauge theory,
provides a non-perturbative gauge-invariant Wilson action of Yang--Mills
theory. We have presented only the basic idea and basic relations in this
paper; we expect many future applications including analytical or numerical
searches for non-trivial RG fixed points in gauge theory. We can also expect
extensions in various directions, such as inclusion of matter fields and
search for a reparametrization-invariant ERG formulation of quantum gravity.
It should be also interesting to clarify a possible relation to the other
gauge-invariant ERG formulations of gauge theory~\cite{Morris:1999px,%
Morris:2000fs,Arnone:2005fb,Wetterich:2016ewc,Wetterich:2017aoy}.

\section*{Acknowledgments}
This work was initiated during the 10th International Conference on Exact
Renormalization Group 2020 (ERG2020). We would like to thank the organizers
and the Yukawa Institute for Theoretical Physics at Kyoto University for
support (YITP-W-20-09). This work was partially supported by Japan Society for
the Promotion of Science (JSPS) Grant-in-Aid for Scientific Research Grant
Numbers JP16H03982 and~JP20H01903.

\appendix

\section{Normalization of the gauge field}
\label{sec:A}
In Sect.~\ref{sec:3}, we have normalized the gauge field~$A_\mu^a(x)$ so that
the rescaled field~$\widetilde{A_\mu^a}(x)\equiv\lambda A_\mu^a(x)$, defined
by~Eq.~\eqref{eq:(3.10)}, has the ordinary gauge
transformation~\eqref{eq:(3.11)}. In fact this is not the only choice of
normalization. We can change the normalization of~$A_\mu^a(x)$ arbitrarily so
that the rescaled field is given by
\begin{equation}
   \widetilde{A_\mu^a}(x)=\lambda z(\tau)A_\mu^a(x).
\end{equation}
Let $S_{z,\tau}[A]$ be the Wilson action of this field. We should then obtain
\begin{align}
   &z(\tau)^n
   \left\langle
   \exp\left[-\frac{1}{2}\int d^Dx\,
   \frac{\delta^2}{\delta A_\mu^a(x)\delta A_\mu^a(x)}
   \right]
   A_{\mu_1}^{a_1}(x_1)\dotsb A_{\mu_n}^{a_n}(x_n)
   \right\rangle_{S_{z,\tau}}
\notag\\
   &=\left\langle
   \exp\left[-\frac{1}{2}\int d^Dx\,
   \frac{\delta^2}{\delta A_\mu^a(x)\delta A_\mu^a(x)}
   \right]
   A_{\mu_1}^{a_1}(x_1)\dotsb A_{\mu_n}^{a_n}(x_n)
   \right\rangle_{S_\tau}.
\end{align}
This implies~\cite{Sonoda:2015bla}
\begin{equation}
   e^{S_{z,\tau}[A]}
   =\exp\left[\frac{1-1/z(\tau)^2}{2}
   \int d^Dx\,
   \frac{\delta^2}{\delta A_\mu^a(x)\delta A_\mu^a(x)}\right]
   \exp\left(S_\tau[z(\tau)A]\right).
\end{equation}
For
\begin{equation}
   z(\tau)=1+\epsilon
\end{equation}
where $\epsilon$~is infinitesimal, we obtain
\begin{align}
   S_{z,\tau}[A]-S_\tau[A]
   &=\epsilon\int d^D x\,
   \left\{
   \left[\frac{\delta S_\tau}{\delta A_\mu^a(x)}
   \frac{\delta S_\tau}{\delta A_\mu^a(x)}
   +\frac{\delta^2S_\tau}{\delta A_\mu^a(x)\delta A_\mu^a(x)}\right]
   +A_\mu^a(x)\frac{\delta S_\tau}{\delta A_\mu^a(x)}
   \right\}
\notag\\
  &\equiv
   -\epsilon\mathcal{N}_\tau[A].
\end{align}
Hence, $S_{z,\tau}$ satisfies the same ERG equation~\eqref{eq:(3.25)}
as~$S_\tau$ except with the addition of
\begin{equation}
   -\frac{dz(\tau)}{d\tau}\mathcal{N}_\tau[A]
\end{equation}
on the right-hand side. We can interpret $-[dz(\tau)]/d\tau$ as the
anomalous dimension of the gauge field.

The marginal operator~$\mathcal{O}_0(p)$~\eqref{eq:(3.41)} that we have found
at the end of~Sect.~\ref{sec:3} is in fact the operator~$\mathcal{N}$; we find
\begin{align}
   &\mathcal{N}^*[A]
\notag\\
   &=-\int d^Dx\,
   \left\{
   \left[
   \frac{\delta S_\tau}{\delta A_\mu^a(x)}
   \frac{\delta S_\tau}{\delta A_\mu^a(x)}
   +\frac{\delta^2 S_\tau}{\delta A_\mu^a(x)\delta A_\mu^a(x)}\right]
   +A_\mu^a(x)\frac{\delta S_\tau}{\delta A_\mu^a(x)}
   \right\}
\notag\\
   &=\int d^Dx\int d^Dy\,A_\mu^a(x)A_\nu^a(y)
   \int_p e^{ip(x-y)}
   (p^2\delta_{\mu\nu}-p_\mu p_\nu)
   \left[-\frac{p^2}{(e^{-2p^2}+p^2)^2}+\frac{1}{e^{-2p^2}+p^2}\right]
\notag\\
   &=\int d^Dx\int d^Dy\,A_\mu^a(x)A_\nu^a(y)
   \int_p e^{ip(x-y)}
   (p^2\delta_{\mu\nu}-p_\mu p_\nu)
   \frac{e^{-2p^2}}{(e^{-2p^2}+p^2)^2}
\notag\\
   &=\mathcal{O}_0.
\end{align}

We believe that the right choice of the anomalous dimension is necessary to
obtain a fixed point of the ERG transformation.

\end{document}